\documentclass[lettersize,journal]{IEEEtran}
\usepackage{amsmath,amsfonts}
\usepackage{algorithmic}
\usepackage{algorithm}
\usepackage{array}
\usepackage[caption=false,font=normalsize,labelfont=sf,textfont=sf]{subfig}
\usepackage{textcomp}
\usepackage{stfloats}
\usepackage{url}
\usepackage{verbatim}
\usepackage{graphicx}
\usepackage{cite}
\usepackage{comment}

\usepackage{algorithm}
\usepackage{array}
\usepackage[caption=false,font=normalsize,labelfont=sf,textfont=sf]{subfig}
\usepackage{textcomp}
\usepackage{stfloats}
\usepackage{url}
\usepackage{verbatim}
\usepackage{cite}

\usepackage{comment}
\usepackage{bm}
\usepackage{xcolor}

\begin{document}

\title{EnThM: \textbf{En}ergy \textbf{Th}eft \textbf{M}itigation in Smart Grids using Hierarchical Verification of Metering Data} 

\author{
Tapadyoti Banerjee, Pabitra Mitra, {\it Senior Member, IEEE}, and Dipanwita Roy Chowdhury, {\it Senior Member, IEEE}
\thanks{Department of Statistics and Data Science, National University of Singapore and Department of Computer Science and Engineering, Indian Institute of Technology Kharagpur}
\thanks{}}

\markboth{}%
{Shell 
\MakeLowercase{\textit{et al.}}: EnThM: \textbf{En}ergy \textbf{Th}eft \textbf{M}itigation in Smart Grids using Hierarchical Verification of Metering Data}

\maketitle

\begin{abstract}
The advent of digital technologies has revolutionized traditional power distribution networks, transforming them into smart grids that are more reliable, efficient, and sustainable. Despite these advancements, electricity theft remains a significant threat to the effective operation of large electrical networks.
To address this issue, we propose EnThM, a lightweight and communication-efficient scheme for real-time mitigation of power theft in smart grid systems. Our approach uses the hierarchical structure of the smart grid infrastructure to verify the authenticity of the metering data at multiple levels of the power distribution network. 
Our work focuses primarily on issues related to cryptographic security.
The verification process involves statistically modeling the cumulative averages of the power usage data and applying rule-based checks on the aggregated power consumption at each level, while accounting for seasonal and daily consumption variations. The proposed method has been tested on benchmark consumption data, yielding high accuracy, efficient implementation, and real-time applicability.  
\end{abstract}


\section{Introduction}\label{Introduction}
Cybersecurity aims to protect systems, networks, and data from cyber threats. As cyberattacks continue to rise worldwide, securing digital systems has become a top priority. Check Point’s 2025 Cyber Security Report ~\cite{CheckPoint} indicates that the average number of weekly cyberattacks on enterprises reached 1673, a 44\% increase over the previous year. This surge is driven by expanding digital footprints and an increasingly sophisticated cybercrime network. Organizations now face advanced, coordinated attacks as well as large-scale social engineering campaigns. Therefore, neglecting cybersecurity can lead to severe consequences, including financial losses, reputational damage, and regulatory penalties.  Cybersecurity encompasses multiple domains, including AI security, application security, cloud security, network security, endpoint security, mobile security, information security, and critical infrastructure security. Among these, critical infrastructure security is particularly important because of its direct impact on public safety, economic stability, and national security. Disruptions to critical infrastructure can have far-reaching and devastating consequences, affecting everything from public health to national defense. In this paper, we focus on the cybersecurity of power grid systems, one of the most prominent examples of critical infrastructure. 

Over the years, power grids have evolved from small, localized systems to large, widely distributed networks, spanning entire nations or even continents. Electricity grids are now recognized as a vital component of critical infrastructure in many countries. However, a significant concern is that much of the existing grid infrastructure was established over a century ago, which is unable to meet the growing demands due to its aging components and the lack of modern technological capabilities~\cite{verdejo2020erratic,smajla2021influence,pereira2015consumer}. This issue is expected to worsen in the coming years, as forecasts predict that global energy demand will triple by 2050, compared to current levels~\cite{chen2021securing}. The energy sector has adopted digital technology at a slower pace than other industries, primarily due to its large scale and the need for high system availability. To maintain the stability and reliability of the electricity supply, the current power system requires urgent and comprehensive reform. These reforms must be cost-effective and compatible with existing infrastructure to ensure an uninterrupted transition. To address this challenge, conventional power systems are being upgraded to smart grids to meet the needs of the contemporary world. A smart grid (SG) enables bilateral flows of electricity and information, resulting in improved energy distribution, increased dependability, and a robust platform for integrating renewable energy sources. 
Although such connectivity improves power distribution and smart monitoring, it also exposes the network to critical cyberthreats, which could cause widespread power outages, compromised data, and financial losses.
As a result, numerous initiatives have been launched to develop communication standards, protocols, and technologies for smart grids~\cite{gungor2011smart}. One such important protocol is IEC 61850~\cite{IEC61850}. However, this did not include any cyber or information security features.
Although the improved protocol IEC 62351 includes some
comprehensive security measures~\cite{IEC62350}, major vulnerabilities persist~\cite{strobel2016novel}.
One of the key weaknesses that remains is the occurrence of unexpected and abrupt power outages, which continue to have a significant impact on consumers and distributors in current power infrastructures. In this paper, our focus is specifically on monitoring and detecting power theft arising from unknown, misallocated, or inaccurate energy flows. 

In a power grid, electricity losses can be broadly classified into two categories: technical losses and non-technical losses. Technical losses occur naturally during the transmission and distribution of electricity, due to the inherent resistance of electrical components, such as cables, overhead lines, and transformers. These losses are unavoidable to some extent, though they can be mitigated through infrastructure upgrades and improved grid design.
On the other hand, non-technical losses arise from factors unrelated to the physical properties of the electrical system. These include electricity theft, faulty metering, and unmetered consumption.  Such losses are often the result of cyberattacks, in which malicious users exploit vulnerabilities in the power system to consume large volumes of electricity without detection. This imposes a considerable load on electrical systems, leading to substantial financial losses and energy waste. 
Non-technical losses can be further broadly categorized as follows:  
\begin{itemize}
\item {\it Energy Theft}: This involves unlawful extraction of electricity from the grid, typically through meter tampering or the manipulation of electrical equipment. 
\item{\it Conveyance Errors}: This occurs when power is used lawfully, but not accurately documented, resulting in unaccounted energy loss within the system.
\item {\it Errors in Unmetered Supplies}: Fluctuations in the electricity consumption of unmetered supplies, which include street lighting, public infrastructure, and advertising displays, can lead to non-technical losses when the estimated usage deviates from the real demand. 
\end{itemize}

Among the aforementioned types of non-technical losses, conveyance errors and errors in unmetered supplies are difficult to account for and are unavoidable to a certain extent. However, energy theft resulting from cyberattacks is a critical area for intervention. 
Specifically, in developing countries unlawful use of electricity can potentially rise up to 30\%~\cite{bhattacharyya2005electricity}, and even in developed countries, such as the United States, electricity theft is estimated to cost the energy industry up to \$6 billion each year~\cite{CIO}.
Consequently, to reduce power outages and help mitigate the impending global energy crisis, it has become imperative to develop efficient, real-time monitoring and detection mechanisms for energy theft within smart grid systems. 
This involves the development of a predictive recognition system designed to monitor and analyze daily electricity consumption data for individual customers. 
Existing approaches sometimes utilize machine learning-based classifiers,  such as Support Vector Machine (SVM) and Artificial Neural Networks, for the detection of electricity theft in smart grids~\cite{nagi2009nontechnical,depuru2011support,depuru2013high,costa2013fraud,guerrero2014improving,zheng2017wide,hasan2019electricity,de2020detection,kocaman2020detection,viegas2018clustering,razavi2019practical,lin2021electricity}. 
However, these methods often  require computationally intensive training and  artificial feature extractions, which increases both space and time complexities, making them unsuitable for high-performance real-time implementation~\cite{kocaman2020detection}. 
A real-time power theft monitoring and detection system with a double metering architecture has been proposed recently in~\cite{zulu2023real}, however, its practical feasibility is limited due to high implementation costs. 

In this paper, we propose EnThM, a novel real-time strategy for mitigation of energy theft by tracking the usage data from the smart meters, specifically focusing on issues related to cryptographic security. 
A smart meter transmits the global intensity data (that is, the magnitude of the power consumed), in short intervals, usually every fifteen to thirty minutes. Our approach involves tracking the cumulative averages of global intensity over a sliding time window of length $T$, initialized at the present time and also at a corresponding point 12 months prior. These averages are computed whenever the measured intensity deviates above or below the baseline current level. The cumulative averages are updated in real time, based on a rate-of-change parameter that is dynamically calculated from the global intensity data itself. Variations in the global intensity are then compared to the evolving cumulative averages, using the cumulative distribution function of the uniform distribution, to assess the legitimacy of the consumption pattern and detect potential electricity theft.  
The following are the highlights of our proposed method:  
\begin{itemize}
    \item Our method for electricity monitoring and theft detection relies on a simple statistical analysis of the smart meter usage data. As a result, it can be implemented efficiently and in real time, enhancing the security of smart grids against cyberattacks. 
    \item We leverage the hierarchical structure of the smart grid infrastructure to enable periodic, key-less authentication of the identities of the consumers. 
    \item The proposed method has been tested on benchmark data, where we also consider variations in electricity consumption patterns due to both seasonal and daily fluctuations, demonstrating high accuracy for identifying electricity theft from smart-meter usage data. 
\end{itemize} 

The remainder of the paper is organized as follows: Section~\ref{BG} introduces the smart grid architecture, discusses various threat models, and provides a brief overview of the properties of uniform distribution. In Section~\ref{Paper_ElECTRODES} we describe our proposed method for monitoring and detecting electricity theft. In Section~\ref{Paper_ExperimentalResults} we present the experimental results of our approach. Finally, concluding remarks are provided in Section~\ref{Paper_Conclusion}.

\section{Background and Preliminaries}\label{BG}
In this section, we describe some of the prior work (Section \ref{PW}), the smart grid architecture (Section \ref{GA}),  the threat model that we have considered in our approach (Section \ref{TM}), and the relevant background on probability distributions (Section \ref{UniformDistribution}). 

\subsection{Prior Work}
\label{PW}
Existing approaches for tackling electricity theft in smart grids often utilize machine learning-based classifiers, such as Support Vector Machine (SVM) and Artificial Neural Networks~\cite{nagi2009nontechnical,depuru2011support,depuru2013high,costa2013fraud,guerrero2014improving,zheng2017wide,hasan2019electricity,de2020detection,kocaman2020detection,viegas2018clustering,razavi2019practical,lin2021electricity}.
Specifically, Nagi et al.~\cite{nagi2009nontechnical} introduce a novel methodology for detecting non-technical losses in power utilities using a Support Vector Machine (SVM). The fraud detection model developed in their study identifies suspicious customers by analyzing anomalies in consumption behavior. Their approach employs data mining techniques, including feature extraction from historical consumption data. The SVM-based model utilizes customer load profiles and other relevant factors to detect behavioral patterns that are strongly correlated with non-technical losses. 
The framework of SVMs is also employed in~\cite{depuru2011support}. This work delineates the estimated energy consumption patterns of various clients, including instances of theft. A dataset of consumer energy usage profiles is constructed using historical data, and SVMs are subsequently trained on data collected from smart meters, which captures diverse manifestations of fraudulent behavior. This data is categorized using predefined rules, and anomalous consumption patterns are identified and aggregated accordingly. In a related work, Depuru et al.~\cite{depuru2013high} explore the potential and importance of High Performance Computing techniques in detecting illegal electricity usage. The authors parallelize the entire customer classification process to enhance detection efficiency. In contrast, Costa et al.~\cite{costa2013fraud} suggested the use of a knowledge discovery in databases approach, leveraging artificial neural networks to classify customers for targeted inspection. 
A knowledge-based system was developed by Guerrero et al.~\cite{guerrero2014improving}, that incorporates the experience of field inspectors and employs a combination of text mining, neural networks, and statistical methods to identify non-technical losses. Information was extracted from sample data and translated into rules, which were then integrated with expert-derived rules from inspectors. Zheng et al.~\cite{zheng2017wide} present a novel electricity theft detection method using a hybrid deep learning model that combines a wide and deep Convolutional Neural Network (CNN). The deep component is designed to effectively detect the non-periodicity of theft-related electricity usage patterns, while capturing the periodicity typical of normal consumption. The wide component complements this by capturing high-level characteristics of the usage data. In a similar vein, Hasan et al.~\cite{hasan2019electricity} designed an electricity theft detection system that integrates a CNN with a Long Short-Term Memory (LSTM) network, enabling both spatial and temporal pattern recognition. Another approach for detecting and identifying energy theft in distribution systems is proposed by De Souza et al.~\cite{de2020detection}, introducing a unique methodology tailored to modern power distribution networks. 
In this method a self-organizing map is employed to cluster users based on similar consumption patterns. For each cluster identified by the SOM, a Multilayer Perceptron Artificial Neural Network is developed to classify consumers as either honest or fraudulent. In a different approach, Kocaman et al.~\cite{kocaman2020detection} developed a deep learning method based on LSTM networks to detect electricity theft. 
Viegas et al.\cite{viegas2018clustering} explored cluster-based techniques, employing Fuzzy C-Means and Fuzzy Gustafson–Kessel algorithms for identifying anomalous consumption behaviors. Similarly, Razavi et al.\cite{razavi2019practical} applied a Genetic Programming algorithm to detect electricity theft in smart grid environments. Lin et al.~\cite{lin2021electricity}, on the other hand, suggested an approach using adaptively tuned Recurrent Neural Networks to enhance detection accuracy. 

Most of the methods mentioned above, however, often require computationally intensive training and  artificial feature extractions, which increases both space and time complexities, making them unsuitable for high-performance real-time implementation~\cite{kocaman2020detection}. 
A real-time power theft monitoring and detection system with a double metering architecture has been proposed recently in~\cite{zulu2023real}, however, its practical feasibility is limited due to high implementation costs. 

\subsection{Grid Architecture}
\label{GA}
\begin{figure*}[]
	\begin{center}
		\includegraphics[scale=0.43]{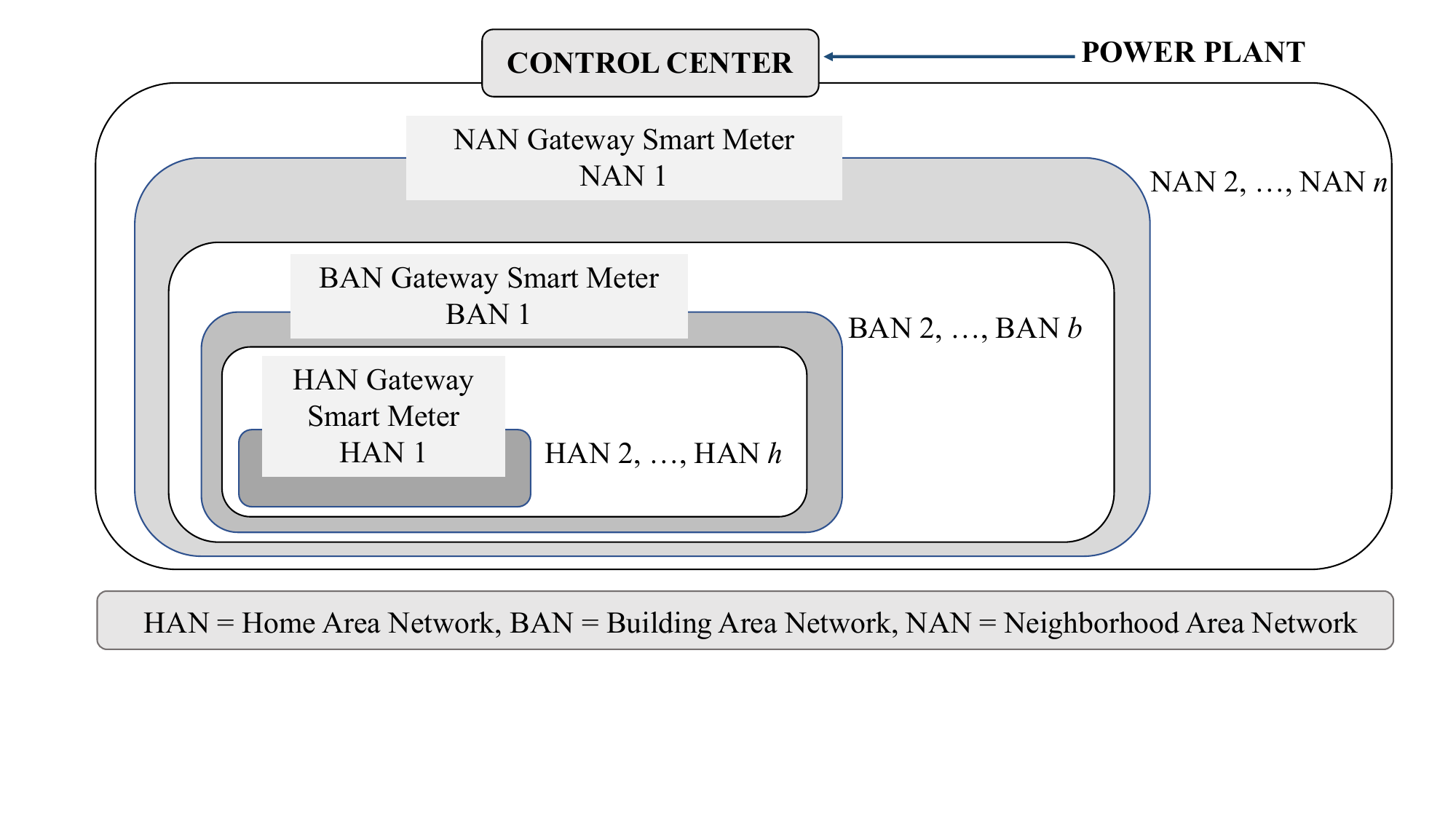}
        \vspace{-0.5in}
		\caption{Hierarchical architecture of Smart Grid (SG)  communication network.} \label{HA_SG}
	\end{center}
\end{figure*}

Figure~\ref{HA_SG} shows a schematic of the Smart Grid (SG) communication framework, which forms the backbone of the power transmission and distribution system~\cite{chim2014prga}.
The power distribution network consists of two main components: the transmission substation of the power plant, and multiple distribution substations. From the perspective of communication, the Smart Grid (SG) topology can be viewed as a hierarchical network structure \cite{fouda2011lightweight,chim2014prga}. 
At the top of the hierarchy is the control center (CC) of a 
Distribution Substation (DS). Each DS comprises of a hierarchical communication network that conists of Neighborhood Area Networks (NAN), Building Area Networks (BAN), and Home Area Networks (HAN). 
For simplicity, we assume that each di
stribution substation serves a single neighborhood area.
For instance,  in Figure~\ref{HA_SG} we show $n$ NANs, each of which is made up of $b$ BANs, each of which in turn has a variety of apartment-based networks (HANs) assigned to them. Smart Meters (SM) are deployed in the SG design to enable automated two-way communication between utility providers and customers. Each smart meter has two interfaces: one for recording power and the other acts as a communication gateway. These meters track key metrics such as electricity usage, voltage, current, and power factor, transmitting data for system monitoring, customer billing, and insights into consumption patterns. This system improves transparency for both consumers and electricity suppliers.

\subsection{Threat Model} 
\label{TM}
We will examine potential attacks on the smart grid structure using the Dolve-Yao threat model~\cite{dolev1983security}. This model posits that attackers can engage in various malicious activities, including eavesdropping on grid communications, intercepting wireless two-way channels, replaying old messages, forging messages, injecting new messages, and the extraction of sensitive information from consumer-to-gateway communications.
The increasing intricacy and widespread interconnectivity of Smart Grids introduce new opportunities for attackers to exploit security vulnerabilities. Inherent weaknesses in the infrastructure can be targeted to compromise control centers, intercept or manipulate smart meter datagrams, disrupt power distribution, and ultimately cause brownouts or even large-scale blackouts. 

\subsection{Statistical Preliminaries}\label{UniformDistribution}
Our power theft detection scheme will use properties of the continuous uniform distribution, which is also sometimes referred to as the rectangular distribution. A distribution of this kind depicts an experiment with an arbitrary result that falls within predetermined parameters. The boundaries are defined by the parameters, $a$ and $b$, which represent the minimum and maximum values of the random variable. The length of the interval is defined by the difference between the boundary values. All intervals of the same length on the support of the distribution are equally likely.  Specifically, the probability density function (PDF) for a continuous uniform distribution on the interval $[a,b]$ is given by:
\begin{equation}
\hspace*{0.4cm} f(x) = {} 
\begin{cases}
1/(b-a) \hspace{1cm} \text{for } a \le x \le b & \\
0 \hspace{2.24cm} \text{for } x < a \text{ or } x > b, & \\
\end{cases}\\
\end{equation}
where the parameters $a$ and $b$ satisfy: $-\infty < a < b < \infty$.  
Hence, the cumulative distribution function (CDF) of the continuous uniform distribution on the interval $[a,b]$ is given by:
\begin{equation}\label{Eq_CDF}
\hspace*{0.4cm} F(x) =  
\begin{cases}
0 \hspace{3.22cm} \text{for } x < a & \\
(x-a)/(b-a) \hspace{1.1cm} \text{for } a \le x \le b & \\
1 \hspace{3.25cm} \text{for } x > b. & \\
\end{cases}\\
\end{equation}
Note that the CDF $F(x)$ is the probability that a random variable with uniform distribution on the interval $[a,b]$ is less than or equal to the value $x$. 

\section{Proposed Methodology}\label{Paper_ElECTRODES}
\begin{figure}[] 
\begin{center}
\hspace{0.75cm}
		\includegraphics[scale=0.43]{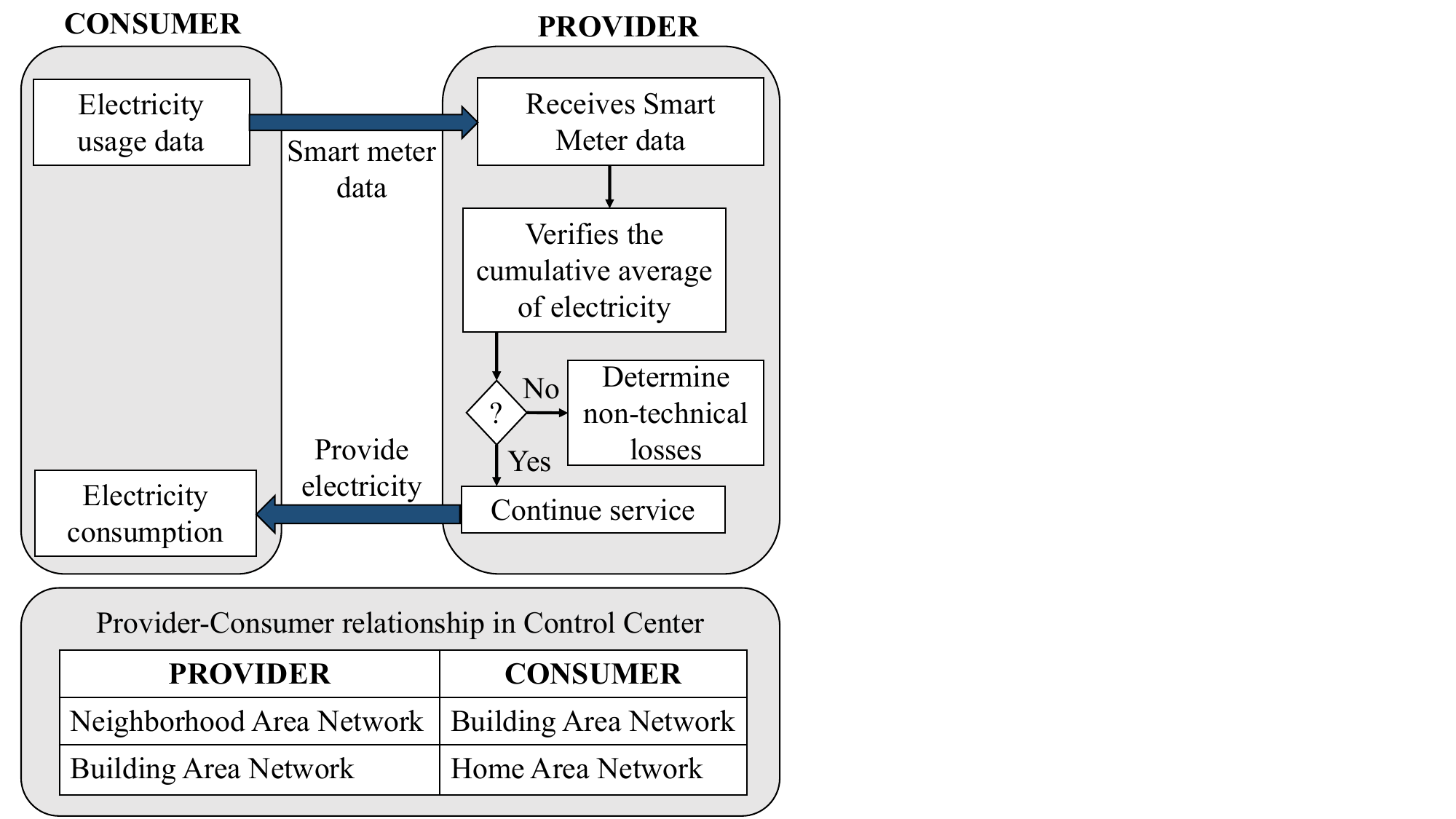}
		\caption{\color{black}{Schematic of the proposed methodology, and the provider-consumer view with respect to smart grid hierarchical architecture.}} 
        \label{BlockDiagram_ELECTRODES}
\end{center}
\end{figure}

The goal of our research is to detect electricity theft by identifying sudden power outages in the smart grid or by detecting non-technical losses. Our proposed method EnThM is a novel real-time, hierarchical verification scheme for energy theft mitigation in Smart Grids. 
Our approach involves analyzing patterns in the cumulative average of electricity usage, based on data periodically transmitted by smart meters. 
The schematic of the proposed approach is given in Figure~\ref{BlockDiagram_ELECTRODES}.
Our method naturally integrates with the hierarchical structure of the Smart Grid (see Figure~\ref{HA_SG}), where the higher-order nodes can execute the algorithm on the electricity consumption data of their immediate lower-order nodes to determine whether there is a power theft. For instance, with reference to Figure~\ref{HA_SG}, each HAN will be verified by its parent BAN, and each BAN will be verified by its corresponding NAN, and each NAN will be verified by the Control Center.

In the practical setting, whenever a new consumer node is added to the grid, it is connected to some producer node, and a first-time authentication takes place to verify its identity~\cite{lakshminarayanan2011authentication}. 
However, since communication between the consumer and producer continues over an extended period, an intial one-time authentication is insufficient to ensure long-term security. The advantage of our verification method is that it can be performed periodically and in real time, each time the smart meter transmits the usage data, ensuring continued safety to the system. It is also worth noting that in the large-scale smart grids, communication routes of the unguided transmission media might be changing as per necessity. However, the communication media does not create any effects on our algorithm, as it primarily deals with the received data and the information stored in the smart meters.

\subsection{Input Parameters} 
Our power theft detection method takes the following parameters as input: 

\begin{itemize}
    \item {\it Time Stamp $(t)$}: Smart meter sends the electricity usage data periodically. Hence, the time stamp $t$  when the data is transmitted is    an important input parameter. 
    
    \item {\it Global Intensity $(GI_t)$}: Another important input parameter is the global intensity of electricity at time $t$. This is the global minute-averaged current intensity (measured in Ampere) as recorded by the smart meter.
    The global intensity at time $t$ will be denoted by $GI_t$. 
\end{itemize}

\subsection{Variable Initialization and Update} 

We denote by ‘$a$’ and ‘$b$’ the lower and upper limits of the current. These are predefined by the service provider, which denotes the maximum upper limit and minimum lower limit of the current at which accuracy is maintained. Alternatively, $a$ is referred to as NULL (denoting no or minimal possible current) and $b$ as $I_{MAX}$, the maximum allowed intensity. Also, denote by $I_b$ the basic current, as defined in IEC 61036
(2000–09)~\cite[Section 3.5.1.1]{IEC61036}. It is the reference value used to define the performance characteristics of a direct connection meter. These parameters are related as follows: 
\begin{equation}
\label{eq:range} 
I_{MAX}= b > I_b > a = NULL . 
\end{equation}  
With the above parameters set, we now compute the cumulative averages of the global intensity $GI_{t_i}$ (depending on whether it is above or below the basic current $I_b$. For this,  denote by $B$ a time span of 1 year (12 months). Hence, given a time instance $t_i$, the notation $t_{i-B}$ will denote the time instance 12 months prior to $i$. Then the cumulative global intensity functions $R_1$ and $R_2$ at time $t_i$ are computed as follows:  
\begin{align}\label{eq:R2}
R_2(t_i) &= \frac{1}{2T+2} \Bigg(\sum_{j=i}^{i+T} GI_{t_j} \bm 1\{GI_{t_j} > I_b\}  \nonumber \\ 
& ~~~~~~~~~~~~~~~~~~ + \sum_{j=i-B}^{i-B+T} GI_{t_j} \bm 1\{GI_{t_j} > I_b \} \Bigg) , 
\end{align} 
and 
\begin{align}\label{eq:R1}
R_1(t_i) &= \frac{1}{2T+2} \Bigg( \sum_{j=i}^{i+T} GI_{t_j} \bm 1\{GI_{t_j} \leq I_b\} \nonumber \\ 
& ~~~~~~~~~~~~~~~~~~ + \sum_{j=i-B}^{i-B+T} GI_{t_j} \bm 1\{GI_{t_j} \leq I_b
\} \Bigg) . 
\end{align} 
Here, $\bm 1\{GI_{t_j} > I_b\}$ denotes the indicator function which is 1 when $GI_{t_j} > I_b$, and 0 otherwise.  $\bm 1\{GI_{t_j} \leq I_b\}$ is defined similarly. 
In other words, $R_2(t_i)$ is  the average of the global intensity, when the global intensity values are higher than $I_b$, over a time slide  window of length $T$ starting at $t_i$ and also over a time slide window of  length $T$ that begins  $B=\text{12 months}$ before $t_i$. Similarly, $R_1(t_i)$ 
is the average of the global intensity, when the global intensity values are lower than $I_b$, over a time slide  window of length $T$ starting at $t_i$ and also over a time slide window of length $T$ that begins $B=\text{12 months}$ before $t_i$. The reason for considering the global intensity values from 12 months prior, in addition to the set of recent values, is to enhance the accuracy of the estimate by considering seasonal variations. By definition of $R_1(t_i)$ and $R_2(t_i)$ and the relation in \eqref{eq:range}, we get 
\begin{equation}
\label{eq:cumulativerange} 
b > R_2(t_i) >  I_b \geq R_1(t_i)  > a . 
\end{equation} 

Next, we continuously update the values of $R_1$ and $R_2$, based on a rate-of-change parameter computed from global intensity data, to obtain the threshold functions which will be used to verify the electricity consumption pattern. The value of the rate of change parameter at time-stamp $t_i$ is denoted by $\alpha$. Technically, the value of $\alpha$ depends on the time-stamp $t_i$, but to simplify the notation we drop the dependence of $t_i$ from $\alpha$. The calculation of $\alpha$ is described later in Algorithm \ref{algorithm:alpha}. 
Then the lower and upper threshold functions $r_1$ and $r_2$ at time $t_i$ are computed as follows: 
$$r_2(t_i) = \left\{\begin{array}{ll}
    b &  \text{ if } t_i = t_0, \\ \\
    (1+\alpha) R_2 (t_i) & 
    \text{ if } (1+\alpha) R_2 (t_i) \le b \\ & \text{ and }  
    b - R_2(t_i)  \ge R_2(t_i) - I_b, \\ \\
    (1-\alpha) R_2 (t_i)
    & \text{ if } (1-\alpha) R_2 (t_i) > I_b \\ 
    & \text{ and } 
    b - R_2(t_i) < R_2(t_i) - I_b , \\ \\
    R_2(t_i) & \text{ if } (1+\alpha) R_2 (t_i) > b, \\ \\
    R_2(t_i) & \text{ if } (1-\alpha) R_2 (t_i) < I_b ; 
\end{array} 
\right.
$$ 
and 
$$r_1(t_i) = \left\{\begin{array}{ll}
    a &  \text{ if } t_i = t_0, \\ \\
    (1-\alpha) R_1 (t_i) & 
    \text{ if } (1-\alpha) R_1 (t_i) \ge a \\ & \text{ and }  
    R_1(t_i) - a \ge I_b - R_1(t_i), \\ \\
    (1+\alpha) R_1 (t_i)
    & \text{ if } (1+\alpha) R_1 (t_i) \le I_b \\ 
    & \text{ and } 
    R_1(t_i) - a < I_b - R_1(t_i), \\ \\
    R_1(t_i) & \text{ if } (1-\alpha) R_1 (t_i) < a, \\ \\
    R_1(t_i) & \text{ if } (1+\alpha) R_1 (t_i) > I_b . 
\end{array} 
\right.
$$ 
From the above definitions and \eqref{eq:cumulativerange} it follows that 
\begin{equation*}
b \ge r_2(t_i) > I_b \ge r_1(t_i) > a . 
\end{equation*}  
The functions $r_1$, $r_2$ are shown in Figures \ref{fig:variables} and \ref{Test1} in the data example. 

\subsection{Calculation of Rate-of-Change Parameter} 
We now describe the calculation of the rate-of-change parameter $\alpha$ at the time-stamp $t_i$. For this, it is also necessary to take into account the values of the global intensity in the current time slide window as well as a time slide window from $B=12$ months prior. The details are given in Algorithm~\ref{algoAlphaELECTRODES}.

\begin{algorithm}
\caption{Rate of change calculation}\label{algoAlphaELECTRODES}
\begin{algorithmic}[1]
 \renewcommand{\algorithmicrequire}{\textbf{Require:}}
 \renewcommand{\algorithmicensure}{\textbf{Ensure:}}
 \REQUIRE Time slide window length ($T$) and the starting timestamp of a window ($t_i$). \\ 
  \FOR {$j = i \text{ to } i+T \text{ and } (i - B) \text{ to } (i - B + T)$}
  \STATE 
  $$n_j = \left\lceil\left(\frac{I_{MAX}- GI_{t_j}}{I_{MAX}} \right) \times 100 \right\rceil.$$
   Express $n_j= 10 \times \alpha_j + \beta_j$, where $0 \leq \alpha_j \leq 10$ and $0 \leq \beta_j \leq 9$ are integers.
  \ENDFOR 
\RETURN $$\alpha = 0.1 \times \mathrm{mode}\{\alpha_{i}, \ldots, \alpha_{i+T}, \alpha_{i-B}, \ldots, \alpha_{i-B+T}\},$$ 
where $\mathrm{mode}$ of a collection of integers is the element that has the highest frequency. In case the mode is not unique, the largest value is chosen. 
 \end{algorithmic} 
 \label{algorithm:alpha}
 \end{algorithm}

To understand the calculation of $\alpha$ described in Algorithm~\ref{algoAlphaELECTRODES} let us consider an example. 

\noindent \textit{Example:} Let us assume that there are 5 samples in the time window from $i$ to $i+T$ with global intensity values as follows: 
\begin{center}
 \{9.600, 9.600, 10.500, 8.600, 7.400\}. 
\end{center}
Also, suppose the time window from ($i-B$) to ($i-B+T$), which also contains 5 more data samples, have global intensity values: 
\begin{center} 
\{11.500, 10.500, 9.600, 9.300, 8.500\}. 
\end{center}
The union of these 2 sets of values is: 
\begin{center}\{9.600, 9.600, 10.500, 8.600, 7.400, 11.500, 10.500, 9.600, 9.300, 8.500\}.
\end{center}
With respect to the standard household electric power consumption, we consider $I_{MAX}$ to be 30 Ampere (Amps)~\cite{WBSEDCL}. Then for each $t_j$ we can compute $n_j$ as in Algorithm~\ref{algoAlphaELECTRODES}. For example, $GI_{t_j} = 9.600$, we have, 
\begin{align*}
n_j & = \left\lceil\left(\frac{I_{MAX}- GI_{t_j}}{I_{MAX}} \right) \times 100 \right \rceil \nonumber \\ 
& = \left\lceil\left(\frac{30- 9.600}{30}\right) \times 100 \right\rceil \nonumber \\ 
& = 68. 
\end{align*}
Similarly, the values of $n_j$ can be calculated for the other values of $GI_{t_j}$. This gives the following set of values for $n_j$: 
\begin{center}
\{68, 68, 65, 72, 76, 62, 65, 68, 69, 72\}.
\end{center} 
Therefore, the corresponding values of $\alpha_j$ are as follows: 
\begin{center}
\{6, 6, 6, 7, 7, 6, 6, 6, 6, 7\}.
\end{center} 
Note that the mode (dominating frequency) in this set is $6$. Hence, $\alpha=0.6$, in this case. \hfill $\Box$

\subsection{ Verification Method } 

We can validate the query global intensity at time stamp $t_{i+T+1}$, using the threshold functions $r_1$ and $r_2$, computed as above. 
Specifically, we check if the global intensity   $GI_{t_{i+T+1}}$ lies in the interval $[r_1(t_{i+T+1}), r_2(t_{i+T+1})]$. In other words, recalling the CDF of the continuous uniform distribution from Equation~\eqref{Eq_CDF}), we say the query global intensity at time stamp $t_{i+T+1}$ is valid when 
\begin{equation}\label{ELECTRODE_CDF}
0 \le \frac{GI_{t_{i+T+1}} - r_1(t_{i+T+1})}{r_2(t_{i+T+1}) - r_1(t_{i+T+1})} \le 1 . 
\end{equation}
Else, we flag the global intensity at the time-stamp $t_{i+T+1}$ as invalid, which prompts the electricity provider to take appropriate action. 

\subsection{ Summarizing the Workflow } 

\begin{figure}[]
\begin{center}
\includegraphics[scale=0.35]{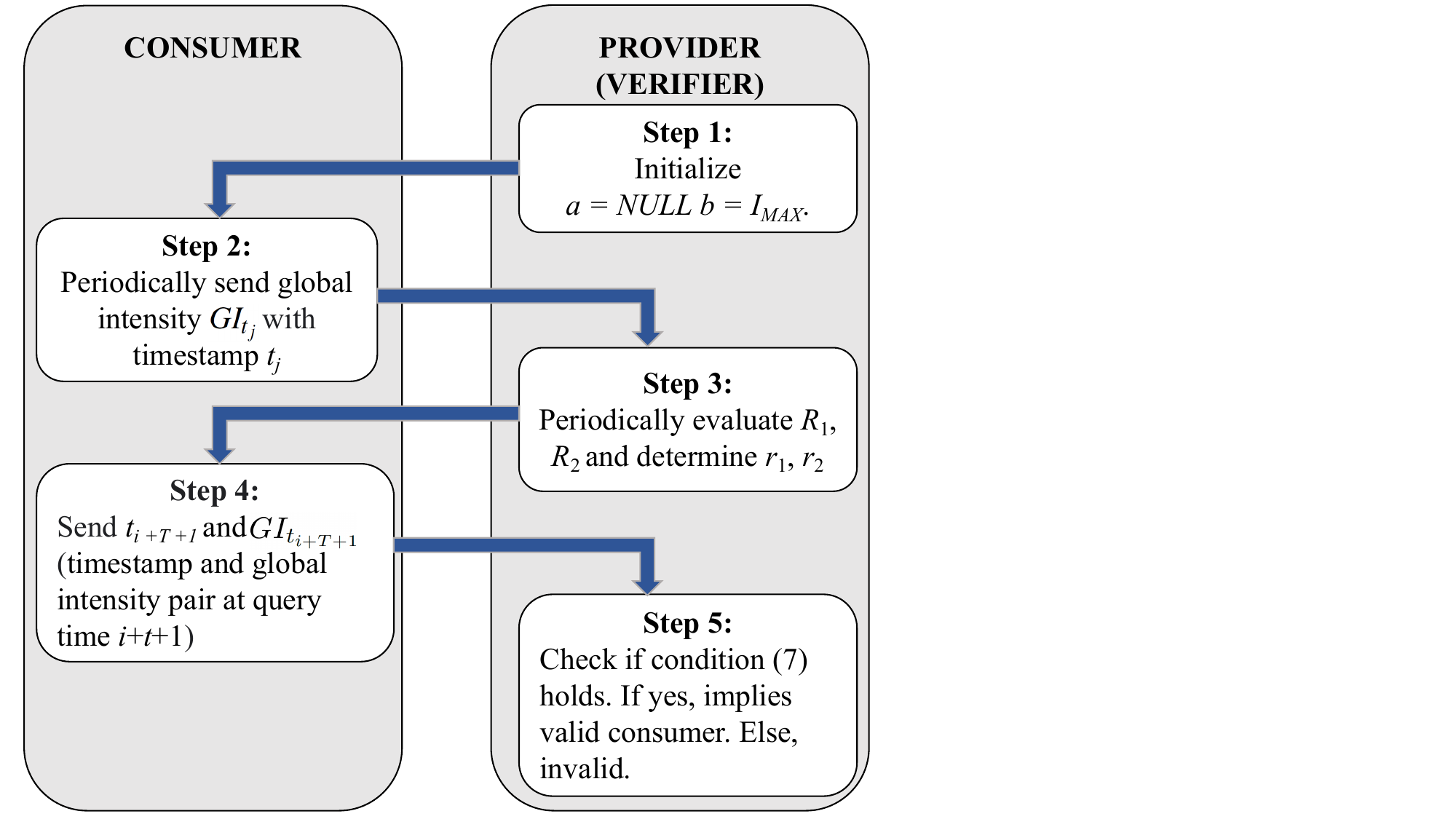}
		\caption{Workflow of the proposed scheme. } \label{Algo}   

\end{center}
\end{figure}

The workflow of our scheme is illustrated in Figure~\ref{Algo}, which highlights the communication flow between the consumer and the provider. The following are the main steps:  
\begin{itemize}
\item Initially, the values of the variables $a$ and $b$ are set by the service provider.  
\item Then the consumer periodically sends the global intensity value $GI_{t_j}$ and its time stamp $t_j$ to the provider.  
\item The provider then calculates the cumulative global intensity functions $R_1$ and $R_2$ and the threshold functions $r_1$ and $r_2$ (as described before).
\item In the query time, the consumer sends the time-stamp $t_{i+T+1}$ and global intensity value $GI_{t_{i+T+1}}$ to the provider. 
\item The provider then verifies the validity of the consumer at time-stamp $t_{i+T+1}$ by checking whether or not condition \eqref{ELECTRODE_CDF} is satisfied.  
\end{itemize}

\section{Experimental Results}\label{Paper_ExperimentalResults}
The details of the experimental setup and results are described in this section. {\color{black}{Note that, since power grid security is part of critical infrastructure security, results from practical testing and comprehensive vulnerability assessments are typically withheld from public access, due to national security concerns and the risk of malicious exploitation. Likewise, data related to critical infrastructure is generally not publicly available, as the exposure of vulnerabilities could lead to catastrophic disruptions of essential services. Governments and commercial organizations responsible for critical infrastructure classify such information as sensitive to protect against threats, including cyberattacks, terrorism, and sabotage. Consequently, we validate our schemes through simulations, benchmark datasets, and FPGA-based testing. } }

\subsection{Data Description} 

For our experiment we use individual household electric power consumption data set provided by the UCI Machine Learning Repository (obtained from Kaggle~\cite{Kaggle}). The dataset contains about 20,75,259 measurements gathered between December 2006 and November 2010 (47 months) with the following 9 attributes: 
\begin{itemize}
\item \textbf{date:} Date formatted as dd/mm/yyyy; 
\item \textbf{time:} Time in the following format: hour, minute, and second (hh:mm:ss); 
\item \textbf{global active power:} Global minute-averaged active power per household (in kilowatts); 
\item \textbf{global reactive power:} Global minute-averaged reactive power per home (in kilowatts); 
\item \textbf{voltage:} Average voltage per minute (in volts) 
\item \textbf{global intensity:} Average household current intensity per minute globally (in ampere); 
\item \textbf{sub metering 1:} Energy sub-metering No. 1 (active energy in watt-hours). This represents the kitchen, which is mostly equipped with microwave, oven, and dishwasher (hot plates are gas-powered rather than electric); 
\item \textbf{sub metering 2:} Energy sub-metering No. 2 (active energy in watt-hours). This includes washing machine, a dryer, a refrigerator, and a light (this is equivalent to the laundry room); 
\item \textbf{sub metering 3:} Energy sub-metering No. 3 (active energy in watt-hours). For this air conditioners and electric water heaters are the equivalents. 
\end{itemize} 
In our analysis, we use the \textbf{date}, \textbf{time} and \textbf{global intensity} attributes.
To introduce anomalies to the data set, we consider Industrial Control System (ICS) Cyber Attack Datasets, generated by Oak Ridge National Laboratories (ORNL)~\cite{ICSdatasets} (which is also available in Kaggle~\cite{Kaggle_PowerSystem}). The dataset includes measurements related to electric transmission system normal, disturbance, control, and cyber attack behaviors. 
In particular, the dataset contains the following types of attack scenarios: 
\begin{itemize} 
\item \textbf{Short-circuit fault:} This is caused by a short in a power line that can occur in various locations along the line. 
\item \textbf{Line maintenance:} Here, one or more relays on a specific line are disabled to allow for maintenance activities on that line.
\item \textbf{Remote tripping command injection attack:} This is an attack that sends a command to a relay, which causes a circuit breaker to open. 
\item \textbf{Relay setting change attack:} Here, relays are configured with a distance protection scheme, and an attacker changes the settings to disable the relay's functionality. As a result, the relay fails to trip in response to a legitimate fault or a valid control command. 
\item \textbf{Data Injection:} This type of attack simulates a valid fault by altering parameters such as current, voltage, and sequence components.  This attack aims to blind the operator and cause a black out. 
\end{itemize} 

\subsection{ Design Rationale }

\begin{figure*}[h!]
	\begin{center}
		\includegraphics[scale=1.015]{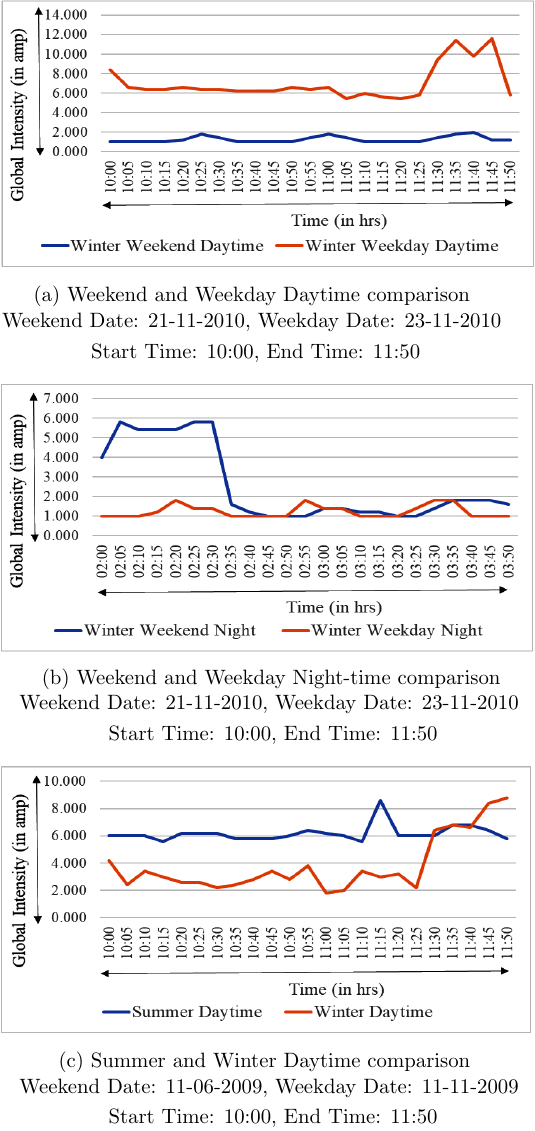}
        \caption{\color{black} Comparing energy consumption patterns. } \label{DesignRationale_3in1}
	\end{center}
\end{figure*}

An important factor to consider when analyzing electricity usage data is seasonal variation. Electricity consumption varies based on the time of day and weather conditions. For example, energy usage patterns differ between day and night, between weekdays and weekends, and also between seasons. 
Figure \ref{DesignRationale_3in1} shows the variations in the load patterns of the dataset in 3 scenarios: (a) comparison of daytime power consumption between weekday and weekend; (b) comparison of nighttime power consumption between weekday and weekend; and (c) comparison of summer versus winter daytime. As expected, power consumption is typically higher during the daytime, when most electrical appliances are in use, compared to night-time. Similarly, energy consumption patterns vary between seasons. In the summer, high-consumption appliances, such as air conditioners, are often used during the day, whereas this usage is significantly lower in the winter. 

The plots in Figure~\ref{DesignRationale_3in1} clearly illustrate that electricity consumption patterns are significantly affected by season and time of day. These variations must be taken into account to accurately detect anomalies in consumption behavior. Hence, in our analysis we implement our monitoring and detection algorithm, that is, construct the threshold functions $r_1$ and $r_2$, separately, depending on the time of the day and the season. 

\subsection{ Results } 
\begin{figure*}[t]
	\begin{center}
		\includegraphics[scale=0.61]{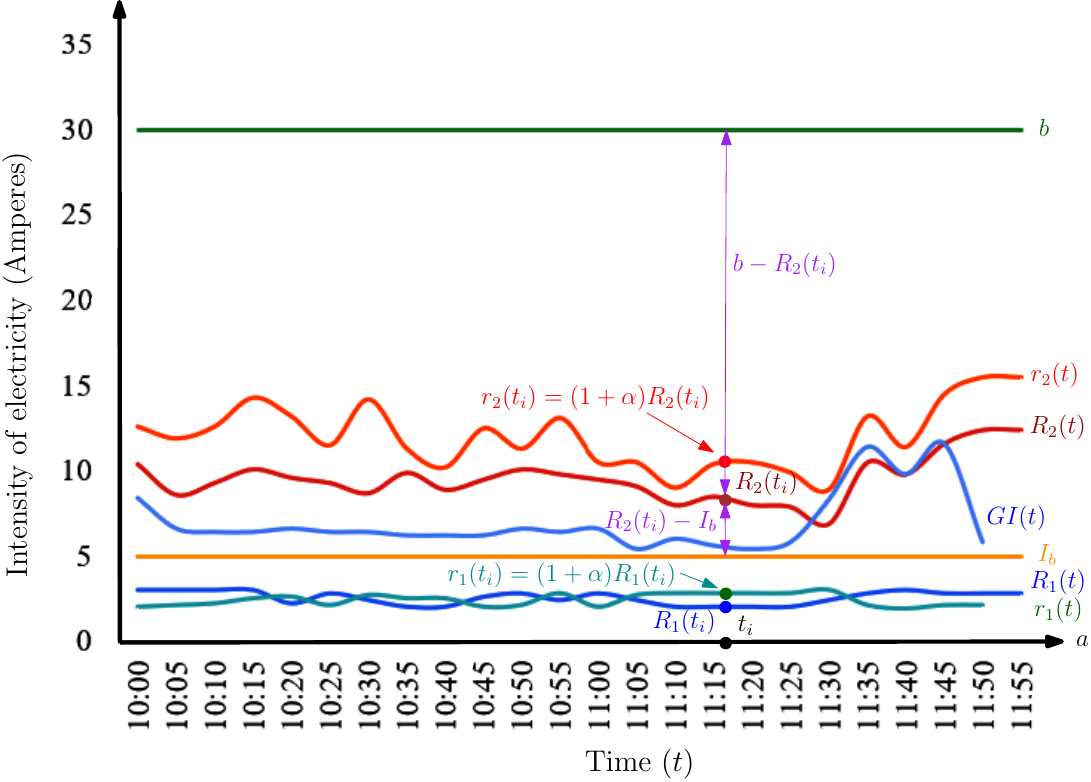}
		\caption{ Calculation of the threshold functions $r_1(t)$ and $r_2(t)$ in the time window 10:00, 21st November, 2010 to 11:50, 21st November, 2010. } \label{fig:variables}
	\end{center} 
\end{figure*}

To illustrate our monitoring and detection method, we consider the time window starting from 10:00 on 21st November, 2010 to 11:50 on 21st November, 2010 (24-hour time format is considered). Figure \ref{fig:variables} shows the global intensity curve $GI(t)$, the cumulative global intensity functions $R_1(t)$ and $R_2(t)$ (computed as in Equations \eqref{eq:R1} and \eqref{eq:R2}), and the threshold functions $r_1(t)$ and $r_2(t)$. The predefined lower and upper limits $a$ and $b$ and the basic current $I_b$ are also shown. Note that in this window, the global intensity curve $GI_t$ lies within the threshold functions $r_1(t)$ and $r_2(t)$, hence no anomaly is reported. 

\begin{figure*}[htb!]
	\begin{center}
		\includegraphics[scale=0.65]{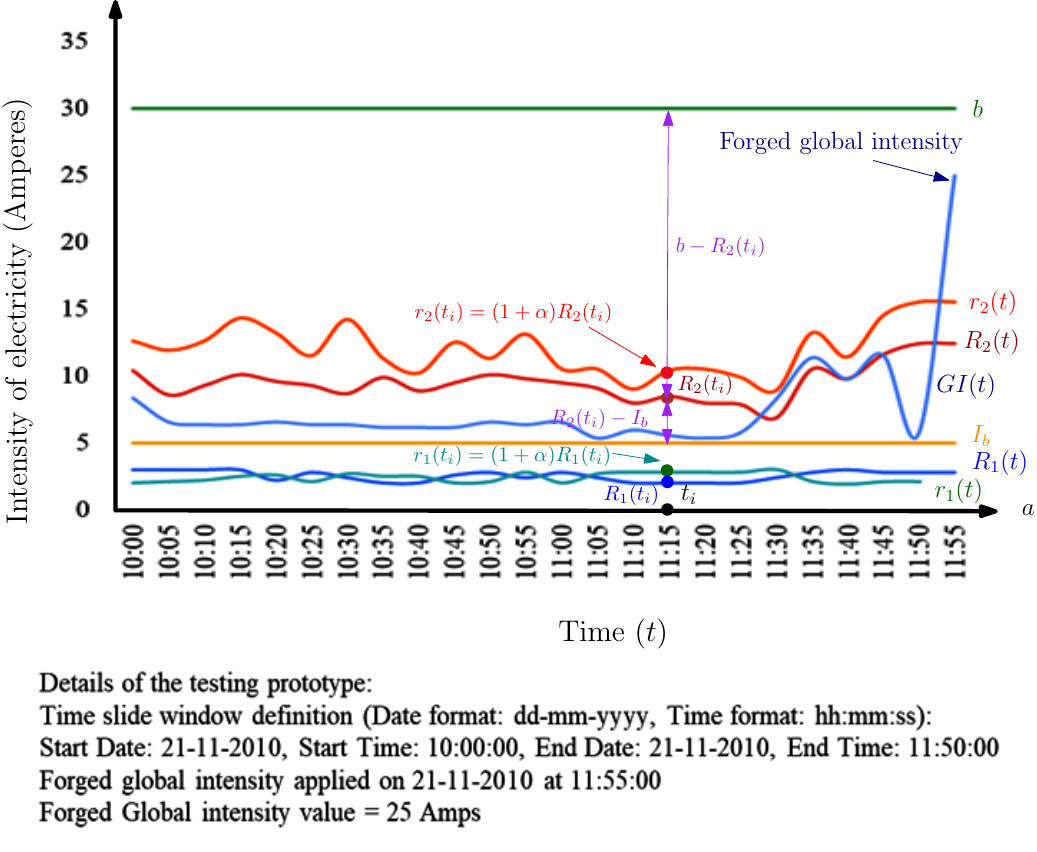}
		\caption{ Anomaly detection at time 11:55, 21st November, 2010. }  \label{Test1}
	\end{center}
\end{figure*}

Next, we apply a forged global intensity at time 11:55 on 21st November, 2010 (this corresponds to the time  $t_{i+T+1}$). Then calculated values of $R_1$, $R_2$, $r_1$, and $r_2$ are 2.800 Amps, 12.400 Amps, 2.100 Amps, and 15.500 Amps,  respectively. (Here, for household electric power consumption, we consider $I_{Max}$ and $I_b$ to be 30 Amps and 5 Amps, respectively~\cite{WBSEDCL}). From this calculate the global intensity value $GI_{t_{i+T+1}}=25 \ \mathrm{Amps}$. Note that, in this case, 
$$\frac{GI_{t_{i+T+1}} - r_1(t_{i+T+1})}{r_2(t_{i+T+1}) - r_1(t_{i+T+1})} = 1.709 > 1 ,$$
which indicates the presence of electricity theft. Hence, our method is able to correctly detect the occurrence of electricity theft in this dataset. Table~\ref{verification} shows the results, and the electricity curve is shown in Figure \ref{Test1}. Note that at the time point 11:55 on 21st November, 2010, the value of the global intensity exceeds the value of the upper threshold function $r_2$ at that point, indicating the presence of an anomaly. 

\begin{table}[h]	
	\centering
	\caption{ Theft verification results }\label{verification}
	\begin{tabular}{|c|c|}
		\hline
		Variables & Values at time $i+T+1$\\
		\hline
		Variable $r_1$ & 2.100 Amps \\
		\hline
		Variable $r_2$ & 15.500 Amps \\
		\hline
		Forged global intensity $GI_{t_{i+T+1}}$  & 25 Amps \\
		\hline
		$\frac{GI_{t_{i+T+1}} - r_1(t_{i+T+1})}{r_2(t_{i+T+1}) - r_1(t_{i+T+1})}$ & $1.709 > 1$ \\
		\hline
	\end{tabular}
\end{table}

Next, we insert forged global intensity values from \cite{ICSdatasets} into the dataset~\cite{Kaggle} multiple times to evaluate the performance of our method. We have injected the faults in the dataset at random positions and executed our proposed method, EnThM, to detect the faults. We have considered 20,75,259 measurements of the dataset~\cite{Kaggle}, within which the number of altered data, or inserted faults, was 500. These 500 faulty data values are taken from the dataset\cite{ICSdatasets}.
Specifically, we take into account the performance of the metrics described here:
\begin{itemize}
\item \textbf{Accuracy:} This is the percentage of correctly identified examples out of all occurrences. Formally, 
\begin{equation*}
\mathrm{Accuracy} = \frac{TP + TN}{TP + FP + TN + FN } . 
\end{equation*}
Here, $TP$ denotes true positives, that is, attack instances correctly identified as attacks; $TN$ denotes true negatives, that is, normal instances correctly identified as normal; $FP$ denotes false positives, that is, normal instances incorrectly identified as attacks; and $FN$ denotes false negatives, that is, attack instances incorrectly identified as normal. 
\item \textbf{True Positive Rate ($TPR$):} This is the fraction of real attacks that are correctly identified: 
\begin{equation}\label{TPFN}
  TPR = \frac{TP}{TP + FN} . 
\end{equation} 
A high true positive rate ($TPR$) indicates the efficiency of a method in detecting malicious activity. 
\item \textbf{False Positive Rate ($FPR$):} This is fraction of typical occurrences that are mistakenly identified as attacks: 
\begin{equation*}
  FPR = \frac{FP}{FP + TN} . 
\end{equation*} 
In order to avoid numerous false alarms, which can make operators less sensitive to alerts and erode their confidence in the detection system, a low $FPR$ is crucial. 
\item \textbf{$F1$ Score:} 
This is defined as 
\begin{equation*}
  F1~\mathrm{Score} = 2 \times \frac{\mathrm{Precision} \times \mathrm{ Recall } }{ \mathrm{Precision} + \mathrm{Recall}}   
\end{equation*} 
where 
  $\mathrm{Precision} = \frac{TP}{TP + FP}$ 
is the proportion of correct positive predictions (true positives) out of all the positive predictions made by the method; 
and 
  $\mathrm{Recall} = \frac{TP}{TP + FN} = TPR$   
is true postive rate (recall \eqref{TPFN}).
Therefore, a high $F1$ Score, which is the harmonic mean of the notions of Precision and Recall, provides a balanced assessment of total performance by indicating that the model performs well in both attack detection and in identifying normal circumstances. 
\end{itemize}

The performance of our proposed method is shown in Table~\ref{DetectionResult}. The results show that our method achieves high Accuracy, TPR, and F1 score as well as low FDP for energy theft detection. Collectively, these metrics highlight effectiveness of our approach in detecting energy theft in smart grid systems.
\begin{table}[h]
	\centering
	\caption{Evaluation Results for detecting Energy Theft}\label{DetectionResult}
	\begin{tabular}{|c|c|}
		\hline
		Accuracy & 99.61\%\\
		\hline
		True Positive Rate &  99.88\%\\
		\hline
		False Positive Rate &  0.25\%\\
		\hline
		F1 Score  &  99.79\%\\
		\hline
	\end{tabular}
\end{table}

\section{Conclusion}\label{Paper_Conclusion}
In this paper, we introduce EnThM, a novel hierarchical real-time verification method for the energy theft mitigation scheme in Smart Grids. The verification process relies on dynamically computing two threshold functions from the smart meter electricity usage data. The observed global intensity values are then compared against these thresholds using the CDF of the continuous uniform distribution, enabling efficient validation of electricity usage at query time. This approach yields a periodic, key-less authentication scheme for threat monitoring and detection that eliminates the need for computationally intensive training or manual feature extraction, making it well-suited for efficient real-time implementation. 

In contrast to existing approaches that focus on predicting the likelihood of future threats, our method is able to detect attacks in real time, as they occur. Once an attack is identified, communication is halted and the provider initiates a verification process.
The method requires only data from a fixed time window $T$, along with corresponding data from up to one year prior. Smart meters collect readings continuously throughout the day and retain this data for 13 months, after which older records are discarded. This simplifies accessing the data and performing the necessary calculations. The proposed method has demonstrated strong performance on benchmark datasets, showcasing its effectiveness in enhancing the security of the Smart Grid infrastructure. 

\bibliographystyle{IEEEtran}
\bibliography{Tapadyoti_PhD.bib}

\end{document}